\documentclass[aip,reprint,amsmath,amssymb,twocolumn,superscriptaddress,showpacs,floatfix]{revtex4-2} 

\usepackage{hyperref}
\usepackage{graphicx} 
\usepackage{dcolumn} 
\usepackage{bm} 
\usepackage{amssymb} 
\usepackage{color} 

\usepackage[utf8]{inputenc}
\usepackage[T1]{fontenc}
\usepackage{mathptmx}
\usepackage{etoolbox}

\usepackage{epsfig}
\usepackage{color}
\usepackage{graphics}

\makeatletter
\def\@email#1#2{%
 \endgroup
 \patchcmd{\titleblock@produce}
  {\frontmatter@RRAPformat}
  {\frontmatter@RRAPformat{\produce@RRAP{*#1\href{mailto:#2}{#2}}}\frontmatter@RRAPformat}
  {}{}
}%
\makeatother

\begin{document}

\preprint{AIP/123-QED}

\title{Glassy dynamics and a growing structural length scale in supercooled nanoparticles.}
\author{Weikai Qi}
\affiliation{Department of Chemistry, University of Saskatchewan, Saskatoon, SK, S7H 0H1, Canada.}

\author{Shreya Tiwary}
\affiliation{Department of Chemistry, University of Saskatchewan, Saskatoon, SK, S7H 0H1, Canada.}

\author{Richard K. Bowles}
\email{richard.bowles@usask.ca}
\affiliation{Department of Chemistry, University of Saskatchewan, Saskatoon, SK, S7H 0H1, Canada.}
\affiliation{Centre for Quantum Topology and its Applications (quanTA), University of Saskatchewan, SK S7N 5E6, Canada.}
\date{\today}

\begin{abstract}

We use molecular dynamics simulation to study the relationship between structure and dynamics in supercooled binary Lennard--Jones nanoparticles over a range of particle sizes. The glass transition temperature of the nanoparticles is found to be significantly lowered relative to the bulk, decreasing as $N^{-1/3}$ with decreasing particle size. This allows the nanoparticles to sample low energy states on the potential energy landscape and we are able to study their relaxation times, measured in terms of the intermediate scattering function, and their structure, measured in terms of locally favoured structures, to low temperatures. Our work shows that the growing relaxation times in the supercooled nanoparticles are coupled with the growth of physical clusters formed from favoured local structures in a way that is well described by the Random First Order Transition entropic droplet model, but with exponents that are dependent on the nanoparticle size.

\end{abstract}

\maketitle

\section{Introduction}
\label{sec:int}
Amorphous nanoparticles exhibit a range of novel properties and play important roles in a variety of natural and engineering processes.~\cite{Chen_Nano_2013}  For example, metallic~\cite{Zhao_Fab_2014} and metal oxide~\cite{Chen_Ultrasmall_2023} glass nanoparticles exhibit catalytic activity for a range of reaction types, while amorphous drug nanoparticles have pharmaceutical applications.~\cite{Jog_Pharmaceutical_2017} Secondary organic aerosols also form highly viscous and and glassy particles at atmospherically relevant temperatures,~\cite{Zobreist_Do_2008,Zobrist_Ultra_2011} where their slow dynamics inhibits water uptake and influences their ice nucleating properties. However,  little is known about the fundamental nature of glass formation in nanoparticles systems, where the finite size and dominant surface properties may change the thermodynamics and kinetics. Many approaches~\cite{Ao_Size_2007,Petters_Predicting_2020,Mahant_Effect_2024} to this problem rely on connections to the size-dependent freezing properties of nanoparticles while current theories of the glass transition in bulk systems are generally independent of freezing. 
Here, we begin to explore the glass forming properties of a simple glass forming nanoparticle system, focusing on the how the nanoparticle size effects the relationship between the slowing dynamics of the system and the growth of a structural length scale. 

Experimentally, the glass transition is a purely kinetic phenomena. As glass forming liquids are cooled, their structural relaxation times increase dramatically over a small temperature range before they become kinetically trapped in an amorphous solid state at the glass transition temperature.~\cite{Berthier_Theoretical_2011,Tarjus_An_2011,Ediger_Perspective_2012,Biroli_Perspective_2013} There is no significant structural change or thermodynamic event associated with the transition, and the properties of the glass are history dependent. However, it has long been argued that the experimental glass transition is the kinetic manifestation of an underlying thermodynamic phenomena, where the observation that structural relaxation times appear to diverge, and an impending violation of the third law of thermodynamics caused by rapidly decreasing liquid entropies,~\cite{Kauzmann_The_1948} suggest the presence of an ideal glass transition located at finite temperature, $T_K$, above absolute zero.


Thermodynamic theories of the glass transition, starting with Adam-Gibbs,~\cite{Adam_On_1965} argue that the supercooled liquid is broken up into a number of metastable states separated by potential energy or free energy barriers. Movement between the states, leading to structural relaxation, then involves the activated motion of the molecules within cooperative rearranging regions (CRR).  As the temperature is lowered, there are fewer accessible metastable states and it takes larger and larger CRR to effect structural relaxation. This gives rise to the suggestion that a diverging relaxation time should be coupled to a diverging structural length scale~\cite{Montanari_On_2006,Montanari_Rigorous_2006,Karmakar_Growing_2009} associated with the size of the CRR.

The structural relaxation times of glass forming liquids exhibit a range of temperature dependent behaviours.~\cite{Angell_Formation_1995} The relaxation times of strong liquids, characterized by network glass formers such as silica, follow an Arrhenius temperature dependence, which suggest the dynamics is controlled by a constant free energy barrier. Fragile liquids, exemplified by many molecular and atomic glass formers, exhibit a super--Arrhenius temperature dependent relaxation time, $\tau$, which implies the free energy barriers to rearrangement, $\Delta$, increase with decreasing temperature, $T$. If the barrier grows due to the increasing size of a structural correlation, $\xi$, then for an activated process we expect $\Delta\sim\xi^{\psi}$, which yields,~\cite{cammarota_Numerical_2009}
\begin{equation}
\ln \tau \sim \frac{\Delta}{T} \sim \frac{\xi^{\psi}}{T}\mbox{.}\\
\label{eq:tau1}
\end{equation}

Unfortunately, the value of the exponent $\psi$ is unknown. One approach is to use the Adam--Gibbs relation, $\ln \tau\sim C/\xi^d s_cT$, where $s_c$ is the configurational entropy per unit volume and $d$ is the dimensionality, in Eq.~\ref{eq:tau1}.  This yields  $\psi=d$, which suggests the structural correlation scales with the number of particles involved in the rearrangement.  Alternatively, the theory of Random First Order Transitions~\cite{Kirkpatrick_Scaling_1989,Bouchaud_On_2004} (RFOT) suggests that the motion between metastable states is driven by a nucleation event where rearrangements then occur when a droplet of size $\xi$ overcomes an activated barrier given by the competition between a generalized surface cost, $\Upsilon \xi^{\theta}$, where $\theta$ is a new exponent, and the entropic driving force, $-Ts_c \xi^d$. Equating these terms and using  Eq.~\ref{eq:tau1} leads to, 
\begin{equation}
\ln\tau\sim(\Upsilon/kT)(\Upsilon/Ts_c)^{\alpha}\sim\left[\frac{A}{T-T_0}\right]^{\alpha}\mbox{,}\\
\label{eq:tau2}
\end{equation}
where $\alpha=\psi/(d-\theta)$. It is also assumed $s_c\sim T-T_K$ and the temperature where $\tau$ diverges, $T_0\approx T_K$, leading to a generalized Vogel--Fulcher--Tamann (VFT) law~\cite{Vogel_Das_1921,Fulcher_Analysis_1925,Tammann_Die_1926} on the right hand side of Eq.~\ref{eq:tau2}. Physical arguments~\cite{Fisher_Nonequilibrium_1988} place bounds on the exponents, $\theta\leq d-1$ and $\theta \leq \psi\leq d-1$. Furthermore, if $\alpha=1$, then the usual VFT law is recovered, as is the Adam-Gibbs relation, although it includes an additional ingredient related to the surface exponent. It was originally suggested $\theta=\psi=3/2$. However, direct measurements find low values for $\psi$ that are outside these bounds.~\cite{cammarota_Numerical_2009,Karmakar_Growing_2009} Identifying the microscopic nature of the structural correlation also remains a significant challenge, but a number of promising candidates have been examined, including point-to-set correlation lengths~\cite{Berthier_Static_2012}, crystal-like order parameter correlations,~\cite{Tanaka_Critical_2010,Tanaka_Bond_2010} polyhedral order correlations~\cite{Xia_The_2015} and favoured local structures.~\cite{Malins_Lifetimes_2013,Ronceray_Favoured_2015,Royall_Role_2015,Hallett_Devil_2020}

An earlier simulation study showed~\cite{Qi_Vapor_2016} that the slow supercooling of liquid nanoparticles allows the system to sample low energy minima on the potential energy surface, similar to those usually only accessible to ultrastable glass films~\cite{Swallen_Organic_2007,Kearns_Influence_2007,Lyubimov_Model_2013} formed through a vapor condensation process that circumvents the liquid. In particular, the extra stability of the nanoparticles resulted from the formation of a large number fraction of favoured local structures (FLS), which are low energy atomic environments connected to slow dynamics. In this work, we study the relationship between glassy dynamics and a growing structural correlation length, measured in terms of the size of physical clusters of FLS, as a function of nanoparticle size using molecular dynamics simulation. The remainder of the paper is organized as follows: Section~\ref{sec:methods} and Section~\ref{sec:res} outline our methods and results respectively. Our discussion is contained in Section~\ref{sec:disc} and our conclusions are summarized in Section~\ref{sec:con}.

\section{Methods}
\label{sec:methods}
We use molecular dynamics simulation to study nanoparticles formed from an 80:20 $AB$ binary Lennard--Jones mixture with Kob--Andersen  (KABLJ) interaction parameters,~\cite{Kob_Testing_1995} $\sigma_{AA}=1.0$, $\sigma_{BB}=1.0$, $\sigma_{AB}=0.8$, $\epsilon_{AA}=1.0$, $\epsilon_{BB}=0.5$, $\epsilon_{AB}=1.5$, and mass, $m=1.0$, for all atoms. The interaction is cut at a distance, $r_c^{\alpha,\beta}=2.5\sigma_{\alpha\beta}$, with $\alpha,\beta\in A,B$, and our results are reported in units reduced with respect to $\sigma_{AA}$ and $\epsilon_{AA}/k_{B}$, where $k_B$ is Boltzmann's constant. The simulations are performed in the canonical ensemble, with the number of particles, $N$, volume, $V$, and $T$, fixed, using the velocity Verlet algorithm and a time step of $0.003$. The simulation box is a cube, with periodic boundaries, and the temperature is maintained using velocity rescaling. Each simulation is started from a random configuration where initial atom--atom distances are greater than $1.5\sigma_{AA}$ and the volume is chosen so that the density, $\rho=N/V=0.15$. The system is initially equilibrated for $10^5$ time steps at $T=0.8$ before it is cooled and condensed into a cluster at a cooling rate, $\gamma=dT/dt$. Two cooling rates are studied, $\gamma=3.3\times 10^{-6}$ and $3.3\times 10^{-3}$. At each $T$ studied, the system is equilibrated again for $4\times10^5$ time steps and the system properties are then measured over the next $2\times 10^6$ time steps. To examine how nanoparticle size influences the structure and dynamics, we study systems in the range $N=200-1500$.

\section{Results}
\label{sec:res}
\begin{figure}
\includegraphics[width=3.2in]{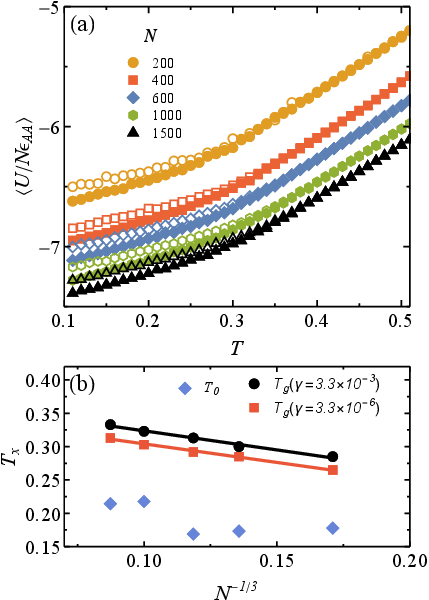}
\caption{(a) Potential energy per particle, $U/N\epsilon_{AA}$, as a function of temperature, $T$, for cooling rates, $\gamma=3.3\times 10^{-3}$(open symbols) and $\gamma=3.3\times 10^{-6}$ (filled symbols) for nanoparticles in the size range $N=200-1500$. (b) Transition temperature, $T_x\in T_g$ or $T_0$ as a function of $N^{-1/3}$. Symbols represent data obtained from simulation and solid lines represent linear fits to the data.}
\label{fig:gt}
\end{figure}

Figure~\ref{fig:gt}a shows the average potential energy per particle, $U/N\epsilon_{AA}$, as a function of $T$, for different nanoparticle sizes and two different cooling rates, $\gamma=3.3\times 10^{-3}$ and $3.3\times 10^{-6}$. At high $T$, the energy is independent of the cooling rate, indicating the liquid nanoparticles are in equilibrium, but they eventually fall out of equilibrium on the time scale of the simulation. This leads to a change in slope in the temperature dependence of the energy and $T_g$ is located at the intersection of linear extrapolations of the high and low $T$ behaviour. Slower cooling rates allow the system to remain in equilibrium longer, causing $T_g$ to move lower. The bulk KABLJ model has been shown to crystallize,~\cite{Pedersen_Phase_2018} but the absence of discontinuities in the energy as a function of temperature indicate there were no nanoparticle freezing events. Figure~\ref{fig:gt}a also shows that the energy of the nanoparticles increases with decreasing $N$ because the small particles have a larger fraction of atoms at the surface which reduces the number of possible neighbour contacts. However, $T_g$ decreases as $N^{-1/3}$ (see Fig.~\ref{fig:gt}b), and extrapolation back to the thermodynamic limit yields a $T_g\approx 0.36$, which is well below the mode coupling temperature for the bulk, $T_{MC}=0.435$, highlighting the degree to which $T_g$ is lowered in these nanoparticle systems. It is also interesting to note that the general phenomenology of the glass transition appears to persist down to nanoparticle sizes as small as $N=200$, where the majority of the atoms are associated with the surface. 

Structural relaxation an a nanoparticle is inhomogeneous, and the dynamics of the atoms at the surface differ from those in the core.~\cite{Hoang_Atomic_2011,Qi_Vapor_2016} Simulation studies~\cite{Berthier_Origin_2017,Sun_Structural_2017} on thin films suggest that the dynamics at the free surface are Arrhenius, but Zhang et al.~\cite{Zhang_Dynamic_2021} found silver nanoparticles exhibit super-Arrhenius surface dynamics that becomes stronger as the nanoparticle size decreases. However, it is also difficult to separate the surface-core dynamics in small nanoparticles, on long time scales, as atoms move between the two regions. Consequently, we measure the dynamics for the whole nanoparticle, which is also consistent with our measurements the nanoparticles energy and glass transition temperature. The dynamics are studied using the self intermediate scattering function, $F_s(q, t)$, measured at the $q$-vector corresponding to the first peak in the structure factor ($q^*\sigma_{AA} = 7.51$).  For all nanoparticle sizes, $F_s(q, t)$ exhibits the same general behavior as a bulk supercooled liquid (see Fig. S1 in Supporting Material). At high $T$, it decays exponentially, but as $T$ is lowered, the initial rapid decay is followed by a shoulder at intermediate times that is characteristic of glassy dynamics. We measure, $\tau$, as the time it takes for $F_s(q^*, t)$ to decay to $1/e$. 

\begin{figure}
\includegraphics[width=3.2in]{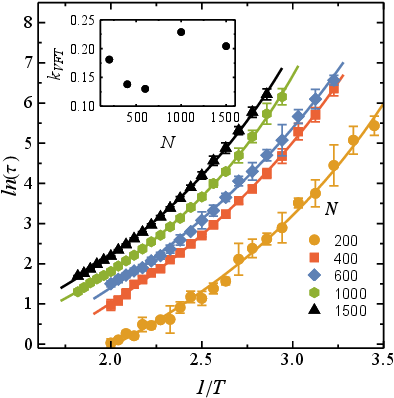}
\caption{Arrhenius plot for nanoparticle relaxation times, $\ln (\tau)$ as a function of $1/T$. The points indicate simulation data and the solid lines represent fits to the data using the VFT equation, Inset: Kinetic  fragility, $k_{VFT}$, as a function of $N$.}
\label{fig:vft}
\end{figure}

It can be difficult to determine the temperature dependence of $\tau$ at low $T$. Mallamace et al.~\cite{Mallamace_Transport_2010} suggested a range of molecular glass formers exhibit a fragile-to-strong liquid crossover at a well defined crossover temperature. However, a more detailed statistical analysis showed that the liquids remain fragile over the entire temperature range studied.~\cite{Chen_On_2012} Fragile-strong crossovers have also been reported for a number of metallic glass forming materials.~\cite{Zhang_Fragile_2010} The lowered $T_g$ of our nanoparticle system allows us to test for the existence of fragile-strong behaviour over an extended temperature range. Furthermore, the inhomogeneous nature of surface-core dynamics may influence the temperature dependence of $\tau$. We perform a statistical analysis of fits to a fragile-strong crossover model for $\tau$, and calculate the slope, $d\ln(\tau)/d(1/T)$, as a function of $T$ (see Supplementary Material for details). These show our data is best described as being fragile at all $T$. Figure~\ref{fig:vft}) shows the data fit to the VFT equation,
\begin{equation}
\tau=\tau_0\exp\left[\frac{1}{k_{VFT}(\frac{T}{T_0}-1)}\right]\mbox{, }\\
\label{eq:vft}
\end{equation}
where $\tau_0$, $k_{VFT}$ and $T_0$ are fit parameters. The temperature dependence of $T_0$ generally follows that of $T_g$ (Fig.~\ref{fig:gt}(b)), decreasing as the nanoparticles become smaller. The parameter $k_{VFT}$ measures the kinetic fragility, with lower values indicating stronger liquid behaviour. Here, $k_{VFT}$, shows no clear trend with an average value of 0.18 (Fig.~\ref{fig:vft}(Inset)).



\begin{figure}
\includegraphics[width=3.2in]{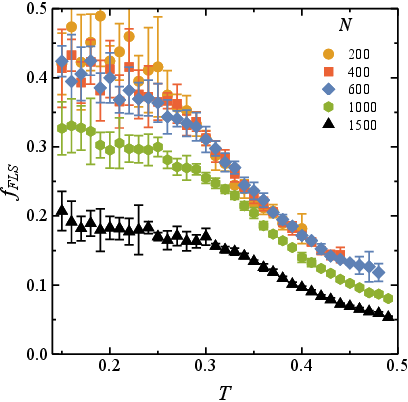}
\caption{Temperature dependence of $f_{FLS}$ for different nanoparticle sizes. }
\label{fig:fls}
\end{figure}

There is strong evidence to suggest that favoured local structures (FLS), which are low energy local environments such as icosahedra, play important roles in the dynamics of supercooled liquids.~\cite{Frank_Supercooling_1952,Tarjus_The_2005,Royall_Direct_2008,Ronceray_The_2011,Malins_Identification_2013,Malins_Lifetimes_2013,Crowther_The_2015,Ronceray_Suppression_2017} In the case of the KABLJ model, the bicapped square antiprism,~\cite{Coslovich_Understanding_2007} which is formed by arranging 10 atoms of any type around a central small $B$--type atom into two five--atom pyramids, has been linked with the energetically stable states of ultrastable glasses~\cite{Lyubimov_Model_2013,Singh_Ultrastable_2013} and dynamically slow trajectories.~\cite{Malins_Lifetimes_2013} Recent studies have also shown that the bicapped square antiprism forms the basis of a dynamical trajectory phase transition between depleted and enriched phases of the FLS,~\cite{Speck_First_2012,Turci_Nonequilibrium_2017,Pinchaipat_Experimental_2017} suggesting that physical clusters or domains of connected FLS could form the basis of a structural length scale. In our simulations, we identify FLS on the basis of Voronoi indices, $\left<n_3,n_4,n_5,n_6\right>$,  that lists the number of faces, $n_i$, with $i$-edges, e.g. triangles, squares, pentagons and hexagons, in the Voronoi polygon surrounding a central atom. The Voro++ code,~\cite{Rycroft_VORO_2009} employing a spherical container to encapsulate the nanoparticle, is used to implement the analysis, and the bicapped square antiprism is identified as having an index $\left<0,2,8,0\right>$. Figure~\ref{fig:fls} shows that the total fraction of atoms that are involved in the $\left<0,2,8,0\right>$ structures, $f_{FLS}$, either as a central atom or one of the ten atoms in the shell, increases as $T$ is lowered for all nanoparticles. It also grows with decreasing nanoparticle size, before reaching an optimal fraction at $N=600$ that is well above the value observed in either the bulk or vapor deposited ultra-stable glass KABLJ model.~\cite{Leoni_Structural_2023}

\begin{figure}
\includegraphics[width=3.2in]{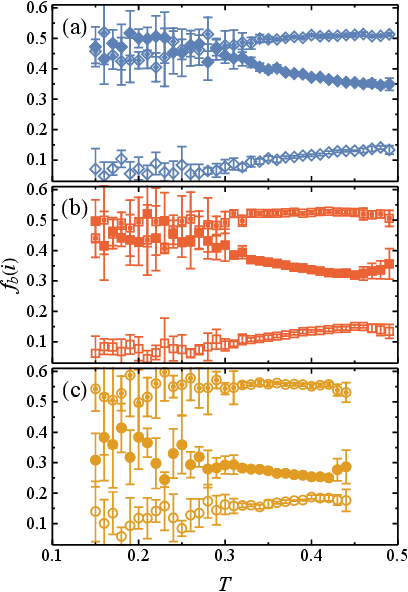}
\caption{The fraction of FLS with the composition $A_{10}B$ ($i=1$ - solid symbols), $A_{9}B_{2}$ ($i=2$ - partially filled symbols) and $A_{8}B_{3}$ ($i=3$ - open symbols)as a function of temperature for nanoparticles sizes (a) $N=600$, (b) $N=400$ and (c) $N=200$.}
\label{fig:flscomp}
\end{figure}

We also follow the composition of the FLS as a function of temperature. Figure~\ref{fig:flscomp} shows the fraction of FLS clusters,$f_b(i)$ containing $i=1,2,3$ B atoms, i.e $A_{11-i}B_{i}$. At high $T$, the $i=2$ FLS dominate, and as $T$ decreases the fraction of the most energetically favoured state,$f_b(1)$, begins to increase at the expense of both  $i=2$ and $i=3$ FLS. For cluster sizes $N\ge 600$ (also see Fig.~S3 in the Supplementary Material)) $f_b(1)\approx f_b(2)$ at and below the glass transition, but $f_b(1)$ decreases with decreasing cluster size. Previous simulations of KALJ nanoparticles~\cite{Chen_Glass_2018} found that the surface was enriched in $A$-type particles which changed the composition of the core and altered the glass forming phase diagram of the system. This suggests that as the nanoparticle size decreases, the core becomes depleted of $A$-type particles, making it more difficult to form the $A_{10}B$ FLS.

\begin{figure}
\includegraphics[width=3.0in]{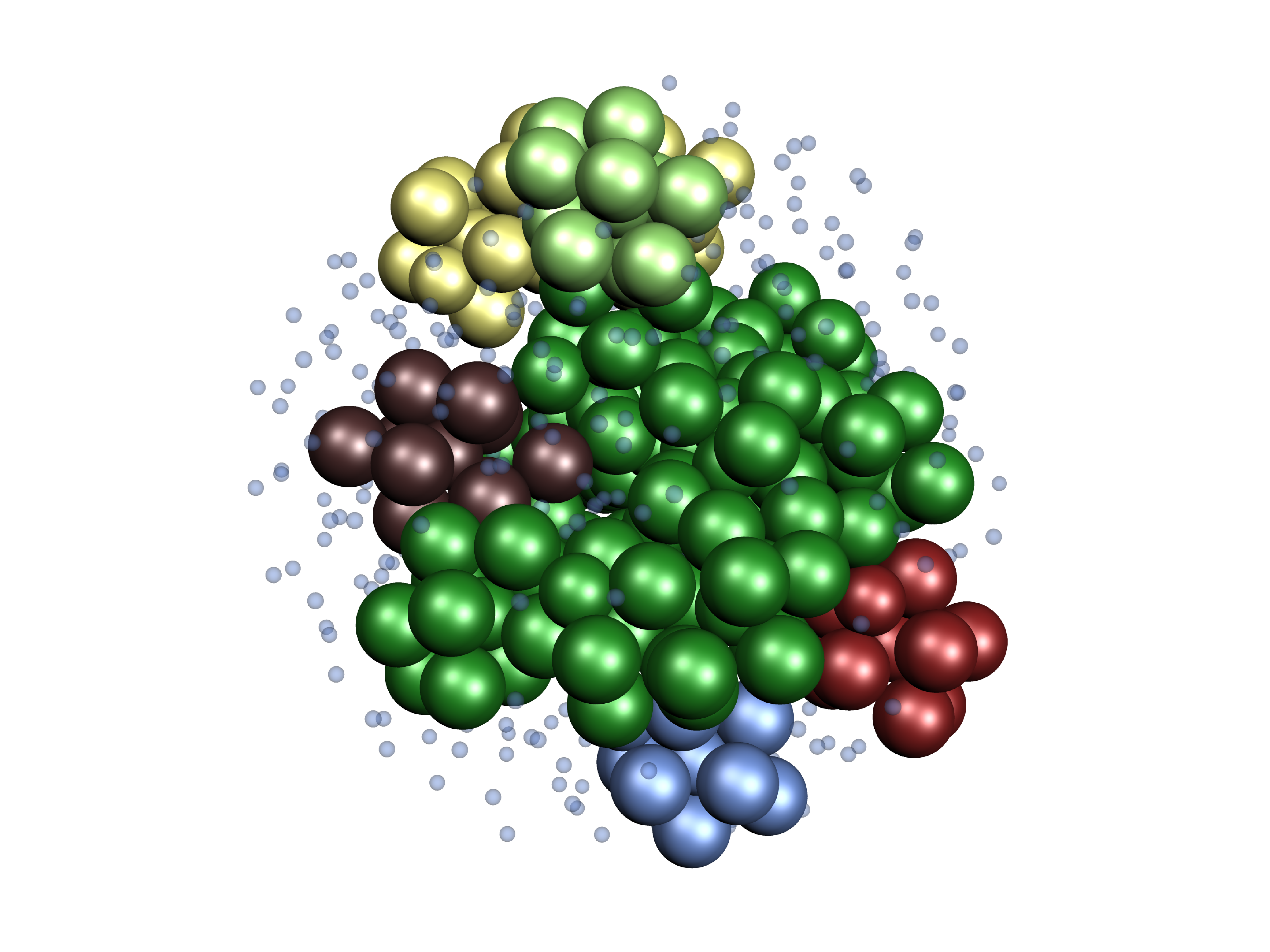}
\caption{A nanoparticle configuration, with $N=600$ and $T=0.3$, showing FLS clusters with different colors. Non-FLS atoms appear as small grey spheres. }
\label{fig:clusters}
\end{figure}

\begin{figure}[t]
\includegraphics[width=3.2in]{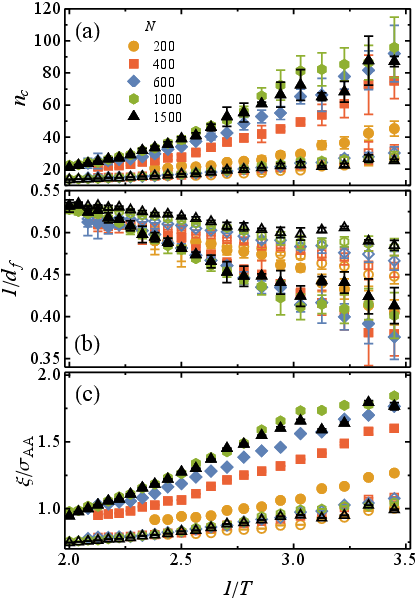}
\caption{FLS cluster properties, (a) cluster size, (b) inverse fractal dimension and (c) structural length scale, averaged over all FLS clusters (open symbols) and the largest FLS cluster (closed symbols).}
\label{fig:cp}
\end{figure}

The FLS within a nanoparticle are divided into physical clusters, where two FLS are considered to be members of the same cluster if they share an atom, and we define a length scale,~\cite{Malins_Lifetimes_2013}
\begin{equation}
\xi=R_g\left(n_c/m\right)^{1/df}\mbox{,}\\
\label{eq:xi}
\end{equation}
where $R_g$ is the radius of gyration of the cluster, $n_c$ is the ensemble average of the cluster size, $m$ is the number of atoms in a cluster and $d_f$ is the fractal dimension, obtained from the scaling exponent, $1/d_f$, of $R_g$ vs $m$. At low $T$, we find that most FLS are collected together in a single large cluster, accompanied by a number of much smaller clusters (see Fig.~\ref{fig:clusters}). Figure~\ref{fig:cp}, shows that the cluster properties, $n_c$, $1/d_f$ and $\xi$, are relatively insensitive to both temperature and nanoparticle size when calculated over all FLS clusters because they are dominated by the presence of a large number of small FLS clusters, often only containing 11 to 20 atoms. However, RFOT theory assumes that only CRR greater than some critical size are able to effect structural relaxation. If we focus on the properties of the largest FLS cluster in each configuration, the properties become much more sensitive, with $n_c$ and $\xi$ growing in a way that is consistent with the super--Arrhenius slow down in the dynamics. The FLS clusters also become more compact compared to those in the bulk,~\cite{Malins_Lifetimes_2013} as indicated by low values of $1/d_f$, possibly because of the finite size of the nanoparticles.

\begin{figure}[t]
\includegraphics[width=3.2in]{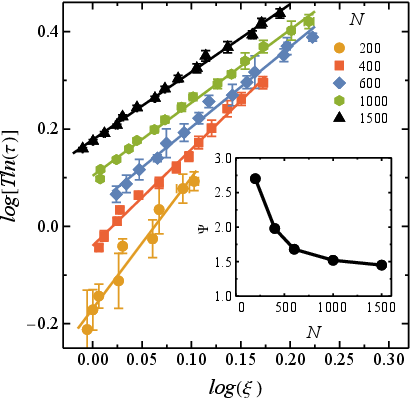}
\caption{$T\log(\ln\tau)$ vs $log(\xi)$ Some curves have been shifted vertically for clarity (unshifted data see Fig. S4 - Supplementary Material). Insert: Nanoparticle size dependence of $\psi$.}
\label{fig:scaling}
\end{figure}

\section{Discssion}
\label{sec:disc}
Random First Order Transition theory has been been successful in describing many of the characteristics of glass forming systems in the bulk.~\cite{Biroli_RFOT_2023} The main goal of this work is to test the applicability of RFOT to nanoparticle systems, where the finite size of the system and the presence of a free surface may play important roles.  According to Eq.~\ref{eq:tau1}, a plot of $T\log(\ln\tau)$ against $\log(\xi)$ should be linear  (Fig.~\ref{fig:scaling}) and the values of $\psi$ (Insert: Fig.~\ref{fig:scaling}) we obtain from the slopes of fits to the data are in the range  $\psi\approx 1.45-2.7$. We do not have a direct measure of $\theta$, but according to the VFT law, $1=\psi/(d-\theta)$, which gives $\theta\approx 1.55-0.3$. In the limit of small nanoparticle size we expect $\psi\rightarrow 3$ and $\theta\rightarrow 0$, as observed, because rearrangements of the core atoms involve the motion of the entire cluster, and the concept of a surface for a CRR, along with the general features of the mean field theory, breakdown.  Here, we note that the application of RFOT to  ultra-stable glass thin films~\cite{Stevenson_On_2008} showed that assuming $\theta=0$ at the film surface leads to a coupling of enhanced dynamics into the film that extends further than a single correlation length before the bulk behaviour is recovered. This suggests that when the nanoparticle is small, all the atoms, even in the core, experience a measure of enhanced dynamics that then leads to the lowered value of $T_g$.

With increasing $N$, the exponents begin to plateau to toward bulk-like values, near but below $\psi=1.5$. Previous estimates~\cite{cammarota_Numerical_2009,Karmakar_Growing_2009} of the scaling exponents obtained using different methods found $\psi\approx 0.3-1.0$, and $\theta\approx 2.0-2.7$. Simulations~\cite{Ozawa_Does_2019,Ortlieb_Royall_2021} of a variety of glass formers have also shown that $\alpha\sim0.5-0.6$. It is not immediately clear why we obtain different exponents and understanding the relationship between different measures of a structural correlation length remains an open question.  In particular, direct measurements of $s_c$ are still necessary to establish the validity of the Adam--Gibbs relation for nanoparticle systems. We also note that our exponents are extracted over a limited range in $\xi$, which makes it more difficult to determine the value of $\psi$, but this is a natural consequence of a length scale based on a physical cluster size, and the finite size of the nanoparticle systems. Nevertheless, our results fall within the expected range and the system size dependence of the exponents is physically reasonable.

Our analysis also suggests that FLS play an important role in the glass forming properties of glass forming nanoparticles. Recent simulations~\cite{Leoni_Structural_2023} have shown that ultra-stable vapor deposited glasses accumulate a large fraction of FLS near the free surface of the film where the enhanced dynamics helps the system sample more stable states. In a nanoparticle, all the atoms are closely coupled to the dynamics of the surface and we find that the fraction of FLS in the nanoparticle system increases as the nanoparticle decreases in size, until an optimal value of $f_{FLS}$ is achieved for sizes $N\le 600$. This may suggest that small nanoparticles systems may sample the bottom of the potential energy landscape, although surface enrichment effects may influence the composition of the FLS and alter the nature of low temperature structure. It is also worth highlighting that we found that the dynamics of the system were more closely related to the size of the largest physical cluster of FLS rather than total fraction of FLS, which may reflect the finite size of the nanoparticles where one large FLS cluster may dominate the dynamics of structural relaxation.

In the bulk KALJ model, the kinetic fragility  increases as function of density and $k_{VFT}\sim 0.22-0.28$ as $\rho=N/V\sim 1.1-1.35$. Our measurements of $k_{VFT}$, obtained through fits to Eq.~\ref{eq:vft}, exhibit significant fluctuations as a function of $N$ and does not have any particular trend. The average value gives $k_{VFT}=0.18$, which is slightly lower than the bulk and may reflect the fact that the nanoparticles sample states that are deeper on the potential energy landscape than those accessible to the bulk. Zhang et al.~\cite{Zhang_Dynamic_2021} showed that small metallic glass nanoparticles exhibit a dynamic fragile-strong crossover as a function of nanoparticle size. In particular, they found that the kinetic fragility of the nanoparticles decreases once the nanoparticle size is below a critical size because the system begins to sample a flat potential energy surface, consistent with a strong liquid. This suggests that our nanoparticle systems may be above the proposed critical size. Nevertheless, our results also suggest there is a lower size limit to which RFOT describes glassy dynamics, related to the size where $\theta\rightarrow 0$ and $\psi\rightarrow 3$, indicating the entire nanoparticle is involved in structural rearrangements.

\section{Conclusions}
\label{sec:con}
In summary, we have demonstrated that a growing structural length scale based on the size of the largest FLS-cluster can account for the super--Arrhenius slow down in the relaxation dynamics in glass forming nanoparticles, and the resulting scaling exponents are well described by RFOT but our work also suggests there is a natural lower limit to the nanoparticle size where the theory would apply. The use of an FLS--cluster based length scale is attractive because physical clusters have been used extensively to understand a range of phenomena, including nucleation in a variety of diverse systems, and may prove useful in understanding features of the proposed dynamical phase between FLS--poor and FLS--rich phases observed in bulk atomic glass formers. Finally, our work shows that the ability of liquid nanoparticles to sample states deep on the potential energy landscape, usually inaccessible to the bulk liquid, make them useful tools in the study of glass structure and dynamics.

\begin{acknowledgments}
We would like to thank NSERC for financial support, and the Digital Research Alliance of Canada for providing computational resources.
\end{acknowledgments}

\section*{Data Availability Statement}
The data that support the findings of this study are available from the corresponding author upon reasonable request.

\section*{Supplementary Material}
Provides additional information regarding the relaxation times, the statistical analysis of the temperature dependence of the relaxations times, and the composition of the $\left<0,2,8,0\right>$ FLS. We also show the unshifted data describing the Random First Order Transition (RFOT) theory exponent scaling.



%




\end{document}